\documentstyle[aps,twocolumn,epsfig]{revtex}
\newcommand{\be}{\begin{eqnarray}}
\newcommand{\ee}{\end{eqnarray}}

\renewcommand{\vec}{\bbox}
\begin{document} 
\title{\bf NON-EQUILIBRIUM REAL-TIME DYNAMICS OF QUANTUM FIELDS: LINEAR
AND NON-LINEAR RELAXATION IN SCALAR AND GAUGE THEORIES}
\author{\bf Daniel Boyanovsky$^{(a)}$, H\'ector J. de Vega$^{(b)}$,
Richard Holman$^{(c)}$,  S. Prem Kumar$^{(c)}$,  Robert
D. Pisarski$^{(d)}$ and Julien Salgado$^{(b)}$} 
\address
{(a) Department of Physics and Astronomy, University of 
Pittsburgh, Pittsburgh  PA. 15260, U.S.A \\ 
(b) LPTHE, Universit\'e Pierre et Marie Curie (Paris VI) et Denis Diderot 
(Paris VII), Tour 16, 1er. \'etage, 4, Place Jussieu, 75252 Paris, Cedex 05, 
France \\  
(c) Department of Physics, Carnegie-Mellon University, Pittsburgh, 
PA. 15213, U.S.A.\\  
(d) Department of Physics, Brookhaven National Laboratory, Upton,
NY 11973, U.S.A.\\
Lecture delivered at TFT98, the Vth International Workshop on Thermal
Field Theories and Their Applications, Ratisbona (Regensburg),
Germany, 10-14 August 1998 }
\date{September 1998}

\maketitle

\begin{abstract} 
The real time evolution of field condensates is solved for small and
large field amplitudes in scalar theories. For small amplitudes, the
quantum  equations of motion for the condensate can be
{\bf linearized} and solved by Laplace transform. The late time evolution
turns to be determined by the singularities in the complex plane
(one-particle poles, two- and multi- particle cuts, Landau cuts for non-zero
initial temperature).
In {\bf hot} scalar electrodynamics, we solve 
the real time evolution of field condensates with soft length scales 
$\sim k^{-1}>(eT)^{-1}$.
We rederive the HTL effective action using the techniques of non-equilibrium 
field theory for small amplitude condensates. We find that transverse gauge
invariant condensates relax as $1/t^2$ and  longitudinal condensates
associated with plasmon (charge density) excitations relax with  
$ 1/[t \; \log^2 t ] $
behavior to asymptotic amplitudes that are determined by the quasiparticle 
poles. 

In the {\bf nonlinear} regime (for large initial energy densities)
we  analyze the dynamics of dissipation
and relaxation in the unbroken symmetry phase of  scalar theory 
after {\bf linear} unstabilities (parametric) are shut-off
by the quantum back-reaction. A new time scale emerges that separates
the linear from the non-linear regimes. This scale  is non-perturbative in
the coupling and initial amplitude. The non-perturbative 
evolution is studied within the context of the $O(N)$ vector model in
the large $N$ limit. 
A combination of numerical analysis and the implementation of a dynamical
renormalization group resummation via multi-time scale analysis reveals
the presence of unstable bands in the nonlinear regime. These are
associated with {\bf power law} growth of quantum fluctuations, that
result in power law  
relaxation and dissipation with {\bf non-universal and
non-perturbative dynamical anomalous exponents}. 
\end{abstract} 
\narrowtext
\section{Introduction}

The dynamics of relaxation of inhomogeneous field configurations out of
equilibrium is an important problem and a common theme in cosmology and high
energy physics.  

In high energy physics the experimental possibility of
studying the chiral and quark-gluon phase transition with high luminosity
hadron colliders and upcoming heavy-ion colliders makes imperative the
understanding of relaxation and transport processes in extreme
environments.

In early cosmology, the energy and time scales during the inflationary
stage of the 
universe calls for an out of equilibrium quantum field
treatment. Moreover, the high energy densities involved ($ \sim 1/g
\sim 10^{12} $) make necessary the use of non-perturbative approaches
as the large $ N $ and Hartree methods\cite{eri97}.

There is currently a great deal of interest in understanding non-perturbative
 real time 
 dynamics in gauge theories at high temperature, both within the realm of heavy
 ion collisions and the study of the quark gluon
 plasma \cite{qgp}-\cite{elze}, as well as the possibility for anomalous baryon
 number violation in the electroweak theory. In both
 situations the dynamics of soft gauge fields with typical length scales 
 $> (gT)^{-1}$ is non-perturbative. 

Their treatment requires a resummation  scheme 
 where one can consistently integrate out the hard scales associated with
 momenta $\approx T$ to obtain an effective theory for the soft
 scales. This is the program of resummation of hard thermal
 loops \cite{resu}.  
 Physically, the hard scale   
represents the typical energy of a particle in the plasma while
the soft scale   
is associated with collective excitations \cite{blaizot2}.

Besides gauge theories,  scalar models as the $ 0(4) $ sigma model are
relevant to study the non-equilibrium
relaxation of an inhomogeneous initial configuration due to quantum and thermal
fluctuations. Collective states of pions as those in connection with
DCC (Disordered Chiral Condensates) can be treated in such a
way\cite{photop,cooper}. 

\bigskip

The main task to solve in this context is the Cauchy problem in
quantum field theory. That is, given the initial quantum description at time
$ t = 0 $, to find it for arbitrary time $ t > 0 $. By quantum
description we mean the initial Fock state in case of pure states and
the density matrix when we have a mixed quantum description.

The dynamics is determined as usual by the Fock space evolution
equations
\begin{eqnarray}
i\frac{\partial|\hat{\Psi}>}{\partial t} &=& H |\hat{\Psi}>\quad
\mbox{pure states} \cr \cr
 i\frac{\partial\hat{\rho}}{\partial
t}&=&[\hat{\rho},H]\quad\mbox{mixed states} \;. 
\end{eqnarray}  
Here $H$ is the second quantized field theoretic Hamiltonian  of the
system and $ \hat{\rho} $ is the density matrix in the Fock space.

The expectation value of any 
operator ${\cal O}$ is given by
\begin{equation} 
\langle{\cal O}\rangle=\text{Tr}[\hat{\rho}(t){\cal O}]\;. 
\end{equation} 
For thermal initial conditions with an initial temperature given by $1/\beta$,
we choose the density matrix at $t=0$ to be 
\begin{equation} \label{hinic}
\rho_i=e^{-\beta H_i}
\end{equation} 
where $ H_i $ is a free (quadratic) Hamiltonian. Notice that $ H_i $
{\em is not} the Hamiltonian of the system for $t>0$. The system thus
evolves out of equilibrium. The above expectation value can be
rewritten easily as a functional integral defined on a complex-time contour. 
The contour has
two branches running forward and backward in time and a third leg along the
imaginary axis stretching to $t=-i\beta$. This is the standard
Schwinger-Keldysh closed time path formulation of non-equilibrium field theory
(see ref. \cite{ctp,nos2,tadpole} for details). Fields defined on the forward
and backward time contours are accompanied with $(+)$ and $(-)$ superscripts
respectively and are to be treated independently. The expectation value of any
string of field operators may be obtained by introducing independent sources on
the forward and backward time contours and taking functional derivatives of the
generating functional with respect to these sources. The imaginary time leg of
the complex time contour does not contribute to the dynamics. 
\section{The Amplitude Expansion in Scalar and Gauge Theory}

Let us consider a quantum scalar field $ \phi(\vec{x},t) $ coupled to
itself and eventually to other fields. The order parameter is then
defined by 
\begin{equation} 
\langle{\phi(\vec{x},t)}\rangle=\text{Tr}[\hat{\rho}(t)\phi(\vec{x},t)]\;. 
\end{equation} 
 For small values of the order parameter, we can linearize the
 evolution equations for the field  $ \phi(\vec{x},t) $ with the
 result\cite{inhomo,tadpole},
\begin{eqnarray} \label{ecmov}
\ddot{\phi}(\vec{x},t)&-&\nabla^2  \phi(\vec{x},t) 
+M^2(T) \; \phi(\vec{x},t) \cr \cr
&+&\int d^3x' \int_0^{t} dt' \; \Sigma_{ret}(\vec{x}-\vec{x'},t-t')\; 
\phi(\vec{x}',t') 
\end{eqnarray}  
Here $ \Sigma_{ret}(\vec{x}-\vec{x'},t-t') $ stands for the retarded
self-energy of the field $ \phi(\vec{x},t) $ and $ $ for the
renormalized mass including temperature corrections. Notice that
eq.(\ref{ecmov}) respects {\it causality} since the field at time $ t
$ only depends on the field values at earlier times $ t' < t $. 

To be more precise, small values of the order parameter for a
self-coupled scalar field of mass $ M $ means that $ \sqrt{\lambda}
\,\phi << M $ where $ \lambda $ is the coupling constant. 

In eq.(\ref{ecmov}), the self-energy kernel  $
\Sigma_{ret}(\vec{x}-\vec{x'},t-t') $ can be computed in perturbation
theory to any finite order. Notice that the self-energy  contains
information about the initial state through the field propagators. 
Namely, perturbative calculations are carried out with the 
non-equilibrium scalar propagators

$${\langle}{\Phi}^{(a)\dagger}(\vec{x},t){\Phi}^{(b)}(\vec{x}, 
t^{\prime}){\rangle}=-i\int {d^3k\over{(2\pi)^3}}\; G_k^{ab}(t,t^\prime) \;
e^{-i\vec{k}\cdot(\vec{x}-\vec{x^\prime})}\;, 
$$
where $(a,b)\;\in\{+,-\}$. 
\begin{eqnarray} \label{propesc}
&&G_k^{++}(t,t^\prime)=G_k^{>}(t,t^{\prime})\Theta(t-t^{\prime}) 
+G_k^{<}(t,t^{\prime})\Theta(t^{\prime}-t)\; , \nonumber
\\ 
&&G_k^{--}(t,t^\prime)= G_k^{>}(t,t^{\prime})\Theta(t^{\prime}-t)+ 
G_k^{<}(t,t^{\prime})\Theta(t-t^{\prime})\;, \nonumber \\
&&G_k^{\pm\mp}(t,t^\prime)=-G_k^{<(>)}(t,t^{\prime})\;, \nonumber\\ 
&&G_k^{>}(t,t^{\prime})=\frac{i}{2\omega_k}\left[ 
(1+n_k)\;e^{-i\omega_k(t-t^\prime)} 
+n_k\;e^{i\omega_k(t-t^\prime)}\right],\nonumber \\ 
&&G_k^{<}(t,t^{\prime})=\frac{i}{2\omega_k}\left[ 
n_k\;e^{-i\omega_k(t-t^\prime)} 
+(1+n_k)\;e^{i\omega_k(t-t^\prime)}\right]\;,\nonumber \\ 
&&\omega_k=\sqrt{\vec{k}^2+m^2}\quad;\;\;\;\;\;\;n_k= 
\frac{1}{e^{\beta\omega_k}-1}\;.  \nonumber
\end{eqnarray}
Here $ \beta \equiv 1/T_i $ and $ T_i $ is the initial temperature as
defined above [eq.(\ref{hinic})].

The explicit expression and the properties of $
\Sigma_{ret}(\vec{x}-\vec{x'},t-t') $ obviously depend on the fields
coupled to $ \phi(\vec{x},t) $. Several cases with scalar and
fermionic fields were discussed in ref.\cite{inhomo,tadpole}. 

Let us consider now scalar quantum electrodynamics (SQED). 
In fact, our ultimate goal is to understand
relaxational processes associated with off-shell effects, such as Landau
damping, in a Non-Abelian gauge theory
\cite{blaizotqgp}. What we do  is to treat the same problem in the
context of scalar quantum electrodynamics (SQED) model\cite{htl}. To leading
order, we expect 
that this should be a good analogue of what happens in the Non-Abelian case; it
also has the advantage that it is simpler to deal with, and as we will see
below, it can be cast from the outset in terms of {\em gauge invariant}
variables. This will eliminate any ambiguities associated with the usual
problem of gauge dependence of off-shell quantities.

We will start with some inhomogeneous field configuration which is
excited in the 
SQED plasma at $t=0$. What we want to do is to follow the time development of
this configuration as it interacts with the hard modes in the plasma, and in
particular, we want to know to determine how it relaxes in time.

We use a gauge invariant formulation of SQED. The SQED Lagrangian is
given by    
\begin{eqnarray} 
&&{\cal{L}}=D_\mu\Phi^\dagger\,D^\mu\Phi-m^2|\Phi|^2-\frac{1}{4}F_{\mu\nu} 
F^{\mu\nu}\; ,\nonumber \\ 
&&D_\mu\Phi=(\partial_\mu-ieA_\mu)\Phi\; . 
\end{eqnarray} 
 
We begin with the identification of the constraints associated with gauge
invariance. The Abelian gauge theory has two first class constraints, namely
Gauss's law and vanishing canonical momentum for $A_0$. What we will do is
project the theory directly onto the physical Hilbert space, defined as usual
as the set of states annihilated by the constraints. 
First, we obtain gauge invariant observables that commute with the first class
constraints and write the Hamiltonian in terms of these. All of the matrix
elements between gauge invariant states (annihilated by first class
constraints) are the same as those that would be obtained by fixing Coulomb
gauge $\vec{\nabla}\cdot\vec{A}=0$. The Hamiltonian in the
physical subspace can be written solely in terms of transverse components and
includes the instantaneous Coulomb interaction.
The instantaneous Coulomb interaction can  be traded for a {\em
gauge invariant} Lagrange multiplier field $A_0(\vec x,t)$ (a non-propagating
field whose canonical momentum is absent from the Hamiltonian) linearly coupled
to the charge density $\rho(\vec{x}, t)$ and obeying the algebraic equation of
motion $\nabla^2 A_0(\vec x,t)=\rho(\vec x,t)$.  This leads to the
following Lagrangian density: 
\begin{eqnarray}  
{\cal L}&=&\partial_\mu\Phi^\dagger\,\partial^\mu\Phi 
+\frac{1}{2}\partial_\mu \vec{A}_T\cdot\partial^\mu\vec{A}_T 
-e\vec{A}_T\cdot\vec{j}_T \cr \cr
&-&e^2\vec{A}_T\cdot\vec{A}_T\; \Phi^\dagger\Phi +
\frac{1}{2}\left(\nabla A_0 \right)^2
\nonumber \\ && + {e^2}A^2_0 \; \Phi^{\dagger}\Phi-i 
e A_0\;\left(\Phi
\dot{\Phi}^{\dagger}-{\Phi}^{\dagger}\dot{\Phi}\right) \; , \nonumber \\
\vec{j}_T=&&i
(\Phi^\dagger\vec{\nabla}_T\Phi-\vec{\nabla}_T\Phi^\dagger~\Phi)\;.
\label{current} 
\end{eqnarray} 
\noindent where $A_T$ is the transverse component of the gauge field. 

In order to provide an initial value problem for studying the relaxational
dynamics of charge density fluctuations we introduce an external source $ {\cal
J}_L(\vec x,t) $ linearly coupled to $A_0$ and study the linear response to
this perturbation. Furthermore, it is convenient to introduce external sources
coupled to the transverse gauge fields to study the linear response of {\em
transverse} gauge field configurations.  These external fields could in
principle play the role of a semiclassical configuration coupled to small
perturbations in a linearized approximation. We thus take,
\begin{equation} \label{fuente}
{\cal L} \rightarrow {\cal L}-{\cal J}_L(\vec x,t) A_0(\vec
x,t)-\vec{{\cal J}}_T(\vec x,t)\cdot \vec{A}_T(\vec x,t) \; . 
\end{equation}  

The relaxational dynamics of our initial inhomogeneous configurations is
clearly an out of equilibrium process, and needs to be treated by an
appropriate formalism\cite{ctp,nos2,tadpole}.

The non-equilibrium SQED Lagrangian is given by 
$$
{\cal L}_{noneq}= {\cal L}\left[\vec A_T^+,\Phi^+,\Phi^{\dagger +},A_0^+\right]
- {\cal L}\left[\vec A_T^-,\Phi^-,\Phi^{\dagger -},A_0^-\right]\;. 
$$
Perturbative calculations are carried out with the scalar Green's
functions given above in eq.(\ref{propesc}) and the non-equilibrium photon
propagators:
$${\langle}{A}^{(a)}_{Ti}(\vec{x},t){A}^{(b)}_{Tj}(\vec{x}, 
t^{\prime}){\rangle}=-i\int {d^3k\over{(2\pi)^3}}\;{\cal G}_{ij}^{ab} 
(k;t,t^\prime)\;e^{-i\vec{k}\cdot(\vec{x}-\vec{x^\prime})}\;, 
$$
\begin{eqnarray} 
&&{\cal G}_{ij}^{++}(k;t,t^\prime)={\cal P}_{ij}(\vec{k}) 
[{\cal G}_k^{>}(t,t^{\prime})\Theta(t-t^{\prime}) 
+{\cal G}_k^{<}(t,t^{\prime})\Theta(t^{\prime}-t) ] \nonumber
\\ 
&&{\cal G}_{ij}^{--}(k;t,t^\prime)= {\cal P}_{ij}(\vec{k}) 
[{\cal G}_k^{>}(t,t^{\prime})\Theta(t^{\prime}-t) 
+{\cal G}_k^{<}(t,t^{\prime})\Theta(t-t^{\prime})] \nonumber
\\ 
&&{\cal G}_{ij}^{\pm\mp}(k;t,t^\prime)=-{\cal P}_{ij}(\vec{k}) 
{\cal G}_k^{<(>)}(t,t^{\prime}) \nonumber \\ 
&&{\cal G}_k^{>}(t,t^{\prime})=\frac{i}{2k}\left[ (1+N_k)e^{-ik(t-t^\prime)} 
+N_ke^{ik(t-t^\prime)}\right]\;,\nonumber \\ 
&&{\cal G}_k^{<}(t,t^{\prime})=\frac{i}{2k}\left[ 
N_k\;e^{-ik(t-t^\prime)} 
+(1+N_k)\;e^{ik(t-t^\prime)}\right],\nonumber\\ 
&&N_k=\frac{1}{e^{\beta k}-1}\;. \nonumber
\end{eqnarray}

Here ${\cal P}_{ij}(\vec{k})$ is the transverse projection operator: 
\begin{equation} 
{\cal P}_{ij}(\vec{k})=\delta_{ij}-\frac{k_ik_j}{k^2}\;.\label{projector} 
\end{equation}  
 
With these tools we are ready to begin our analysis of non-equilibrium
SQED.

The field evolution equations in SQED for the transverse gauge field
take then the form
\begin{eqnarray}\label{ecmovA}
&&{\ddot A}_i^T(\vec{x},t)-\nabla^2 A_i^T(\vec{x},t)  \cr \cr
&&+ \int d^3x' \int_0^{t} dt' \; \Sigma_{ret}(\vec{x}-\vec{x'},t-t')_{ij}\; 
 A_j^T(\vec{x}',t')=0 
\end{eqnarray} 
where $ 1 \leq i, j \leq 3 $ and $
\Sigma_{ret}(\vec{x}-\vec{x'},t-t')_{ij} $ stands for the transverse photon
vacuumm polarization. After calculation we find \cite{htl},

\begin{eqnarray} \label{intSi}
&&\Sigma_{\vec k}(s)_{ij} = \delta_{ij} {e^2 \, T^2 \over 6}
-2e^2\int \frac{d^3p}{(2\pi)^3 
\omega_p\omega_{k+p}}\cr \cr
&&\left[(1+n_p+n_{p+k})\frac{\omega_{k+p}+\omega_p} 
{s^2+(\omega_{k+p}+\omega_p)^2}+\right. \\\nonumber  
&&\left. 
+(n_p-n_{k+p})\frac{\omega_{k+p}-\omega_p} 
{s^2+(\omega_{k+p}-\omega_p)^2}\right] 
p_{Ti}\;p_{Tj}\;
\end{eqnarray} 
In the hard thermal loop limit (HTL) $ T >> k, \; T >>m $ this yields,
\begin{equation} \label{sightl}
\Sigma_{\vec k}(s)_{ij} =
\frac{e^2T^2}{12\pi}\int d\Omega\;\hat{p}_{Ti}\;\hat{p}_{Tj}\;
\frac{s}{s+i\hat{p}\cdot{\vec k}}\;
\end{equation}  
We see that the magnetic mass identically vanishes, as expected.
Introducing the 
lightlike vector $\hat{\cal E}=(1,-\hat{p})$ and recognizing that the 
$s$-variable may be replaced by $\partial_t$ in the time domain, we
obtained the  HTL effective Lagrangian for the transverse
electromagnetic field\cite{htl},  
$$
\delta {\cal L}_{trans}=-\frac{e^2T^2}{6}\!\!\int \frac{d\Omega}{4\pi}
A_{Ti}(\vec x,t)\hat{p}_{Ti} 
\left[\frac{\partial_0}{\hat{\cal E}\cdot\partial}\right]\hat{p}_{Tj} \;
A_{Tj}(\vec x,t)
$$
Therefore, the hard thermal loops are resummed by this method in a
very straightforward way. 

We got analogously for the longitudinal part of the gauge field\cite{htl},
$$
\delta {\cal L}_{long}= 
-\frac{e^2T^2}{6}\!\! \left[A_0^2(\vec x, t)+
\int\frac{d\Omega}{4\pi}A_0(\vec x,t)\frac{\partial_0} 
{\hat {\cal E}.\partial}A_0(\vec x,t)\right] 
$$
which is the known result and displays explicitly the
Debye screening mass of charge density fluctuations.

\section{Exact field time evolution by Laplace transform}

The field evolution equations (\ref{ecmov}) and (\ref{ecmovA}) in the amplitude
approximation can be solved exactly by Fourier transform in space 
and Laplace transform in time \cite{htl,inhomo,tadpole}. 
For the scalar case we define,

 \begin{eqnarray}\label{trafos}
\delta({\vec k},t) &=& 
\int\!\!{{d^3 x}\over{(2\pi)^3}} 
\;e^{-i \vec k \cdot \vec x}\, \phi({\vec x},t) \, ,\,
{\tilde\phi}_{\vec k}(s) 
 =  \int_0^{\infty}\!\! dt \; e^{-st}\;\delta({\vec k},t)  \cr \cr
\Sigma_{\vec k}(s) & = & \int_0^{\infty}\!\! dt \; e^{-st}
\int{{d^3 x}\over{(2\pi)^3}} 
\;e^{i \vec p \cdot \vec x }\;\Sigma(\vec x,t) \;. 
\end{eqnarray}
Notice that the Laplace transform variable $ s $ corresponds to $ i $
times the frequency $ \omega $. The evolution equations (\ref{ecmov})
can now be solved exactly,

\begin{equation}\label{solu}
{\tilde \phi}_{\vec k}(s) =\frac{s\; \delta({\vec k},0) +  {\dot
\delta}({\vec k},0)} 
{s^2+k^2 + M^2(T)  + \Sigma_{\vec k}(s)} \;, \label{lapla}
\end{equation}

We have now to perform the inverse Fourier and Laplace transforms:
\begin{equation}\label{antit}
\phi({\vec x},t) = \int d^3 k \;e^{i \vec k \cdot \vec x}\,
\int_{\Gamma} e^{st} \; {\tilde\phi}_{\vec k}(s) {ds \over 2\pi i}
\end{equation}

where $ \Gamma $ is a contour running from $ -i \infty $ to $ +i
\infty $ paralell to the imaginary $ s $ axis and with a small but
nonzero real part. As it is obvious the singularities of 
eq.(\ref{solu}) are crucial for the behaviour of $ \phi({\vec
x},t) $. The zeroes of the denominator are specially important.

In order to evaluate (\ref{antit}) we deform the contour $ \Gamma $ to
the left. Pole contributions are explicitly computed by the residue
theorem. Cut contributions become real integrals of the corresponding
cut discontinuities (see \cite{htl,inhomo,tadpole}). 

In SQED we analogously obtain for the tranverse gauge field,
 \begin{eqnarray}\label{antisqed}
A^T_{i \, \vec{k}}(t)&=& \int_{\Gamma} e^{st} \;  {ds \over 2\pi i}
\frac{s\; A^T_{i \, \vec{k}}(0) +  {\dot  A}^T_{i \, \vec{k}}(0) 
+ \tilde{\cal J}_{Ti}(\vec k,s) }
{s^2+k^2   + \Sigma^T_{\vec k}(s)}
\end{eqnarray}
Here $ \Sigma^T_{\vec k}(s) $ stands for the transverse vacuumm
polarization. 

There are two kinds of cuts in $ \Sigma^T_{\vec k}(s) $  : \begin{itemize}
\item{ the two particle production cuts from $s=\pm 
i(m+\omega_k)$ to $\pm i\infty$ where $m$ is the mass of the scalar.}
\item{ the thermal (Landau) cut running from $-ik$ to $+ik$ }
\end{itemize}
It must be stressed that the Landau cut discontinuity is of order $
T^2 $ whereas the production cut discontinuity is of order $ \ln T $.
Hence, only the  Landau cut will be consider in the HTL limit.

The explicit form for $ \Sigma^t_{\vec k}(i\omega+0^+) $ follows from 
eq.(\ref{intSi}). In the HTL limit we find (see for example \cite{htl})  
\begin{eqnarray} \label{tbelowcut} 
&&\Sigma^t_{\vec k}(i\omega+0^+)=\frac{e^2T^2}{12} 
\left[2\frac{\omega^2}{k^2}+\frac{\omega}{k} 
\left(1-\frac{\omega^2}{k^2} \right) 
\ln\left|\frac{k+\omega}{k-\omega}\right|\right] \cr \cr
&&+i\frac{e^2T^2\pi}{12}\frac{\omega}{k} 
\left(1-\frac{\omega^2}{k^2}\right)\Theta(k^2-\omega^2)=
\Sigma^t_k(\omega)_R+i\,\Sigma^t_k(\omega)_I \nonumber
\end{eqnarray} 

In addition to the branch cut singularities, the retarded propagator 
 has isolated poles corresponding to the quasiparticle excitations which
can propagate in the plasma. In the HTL limit they are at
$ s=\pm i\omega_P\simeq\pm i\frac{eT}{3} $.

After deforming the contour $ \Gamma $ we obtain \cite{htl}
\begin{eqnarray} 
{\cal A}_{Ti}(\vec k,t)={\cal A}_{Ti}^{pole}(\vec k,t) 
+{\cal A}_{Ti}^{cut}(\vec k,t)\; ,\label{tpolecut} 
\end{eqnarray}

The contributions from the  
quasiparticle poles add up to give a purely oscillatory behavior in time. The
residues at the poles give rise to a  $T$-dependent wave function
renormalization. We thus obtain,  
\begin{eqnarray} 
&&{\cal A}_{Ti}^{pole}(\vec k,t)=
Z^t[T]\! \! \left[ {\cal A}_{Ti}(\vec k,0)   \cos\omega_Pt + \dot{\cal
A}_{Ti}(\vec k,0) \frac{\sin\omega_Pt}{\omega_P} \right] ,\cr \cr
&&Z^t[T]=\left[1-\frac{\partial\Sigma^t(i\omega)}{\partial\omega^2}\right] 
_{\omega=\omega_P\simeq eT/3}^{-1} \label{zetatrans} \; .
\end{eqnarray}

Here $Z^t(T)$ is the temperature dependent wave function 
renormalization defined  
on-shell at the quasiparticle pole whose leading HTL contribution is obtained
from the self-energy (\ref{tbelowcut}). The cut contribution is given by 
 
\begin{eqnarray}
&&{\cal A}_{Ti}^{cut}(\vec k,t)=
\frac{2}{\pi}\; 
\int_0^k d\omega\; \Sigma^t_I(i\omega+0^+)\cr \cr
&&{{\cal A}_{Ti}(\vec
k,0)\; \omega \; \cos(\omega t)+ 
\dot{\cal A}_{Ti}(\vec k,0) \; \sin(\omega t) \over
\left[ \omega^2-k^2-\Sigma^t_k(\omega)_R\right]^2+\left[
\Sigma^t_k(\omega)_I\right]^2}\; .\label{tcutonly}
\end{eqnarray} 
Evaluating Eqs.(\ref{tpolecut}),(\ref{zetatrans}) and (\ref{tcutonly})
at $t=0$ we obtain an important sum rule, 
\begin{eqnarray} 
Z^t[T]+\frac{2}{\pi}\int_0^kd\omega\frac{\omega\Sigma^t_k(\omega)_I} 
{\left[\omega^2-k^2-\Sigma^t_k(\omega)_R\right]^2+
\left[\Sigma^t_k(\omega)_I\right]^2}=1
\label{sumrule} 
\end{eqnarray}
The integral over the cut (\ref{tcutonly}) cannot be evaluated in closed form
but its long time asymptotics is dominated by the end-point
contribution form $ \omega = k $. We find in the long time
limit $ t>>1/k $
\begin{eqnarray}\label{extremo}
&&{\cal A}_{Ti}^{cut}(\vec k,t)\buildrel{t \to \infty}\over=-
 \frac{12}{e^2T^2}\left\{{\cal A}_{Ti}(\vec
 k,0)\;\frac{\cos(kt)}{t^2}\right. \cr\cr &&+ \left. \dot{\cal A}_{Ti}(\vec
 k,0)\;\frac{\sin(kt)}{kt^2}\right\}
\left[1+{\cal O}\left(\frac{1}{t}\right)\right]\quad .
\end{eqnarray} 

We want to stress that the real time dynamics of the condensate is
completely determined by the analytic structure of the retarded propagator and
the {\bf global structure} of the spectral density in the $s$-plane.
 
The second important point to note is that the long time behaviour is a {\bf
power law} $\sim t^{-2}$ (times oscillations) and {\bf not an exponential
decay}. 

This means that Landau damping effects {\bf cannot} be reproduced by
phenomenological `viscous' terms of the type $\sim\Gamma \; \frac{d}{dt}$
neither at long nor at short times. The failure of such a phenomenologically
motivated ansatz was already noticed at zero temperature in different contexts
in ref.\cite{nos2}.  We stress that such a description not only fails to
reproduce the power law behaviour but in fact ignores {\it all the non-local}
physics of Landau damping which is so clearly encoded in the Hard Thermal Loop
kernels.

A very similar analysis yields the longitudinal component of the gauge field
in the case of an impulsive source ${\cal J}_L(\vec x,t) =
\delta^3(\vec x) \delta(t)$. The case of a more complicated source can be 
obtained by convolution. In this case we obtain \cite{htl}
\begin{equation}
{\tilde{\cal A}}_0(\vec k,s)=\frac{1}{k^2+\Sigma^l(s)}
\end{equation}
where  in the HTL limit
\begin{equation} 
\Sigma^l(s)=\frac{e^2T^2}{3}-\frac{e^2T^2}{6} 
\frac{s}{ik}\;\ln\left(\frac{is-k}{is+k}
\right)\; .\label{htlsigmal} 
\end{equation} 
Thus the longitudinal self-energy along the imaginary axis in the $s$-plane, 
when approaching from the right is obtained as before to be: 
\begin{eqnarray} \label{lbelowcut} 
\Sigma^l(i\omega+0^+)&=&\frac{e^2T^2}{3}\left[ 
1-\frac{\omega}{2k}\ln\left|\frac{k+\omega}{k-\omega}\right|
\right]\cr \cr
&-&i\frac{e^2T^2\pi}{6}\frac{\omega}{k}\;
\Theta(k^2-\omega^2)\; . 
\end{eqnarray} 
The location of the longitudinal quasiparticle or plasmon poles is
given again for soft external momenta $ (k<<eT) $ by
$s=\pm i\omega_0\simeq \pm ieT/3$. The real time dependence of the 
longitudinal condensate is then found by inverting the transform through 
\begin{eqnarray}
{\cal A}_0(\vec k,t)=\int_{c-i\infty}^{c+i\infty}\frac{ds}{2\pi i}\;
e^{st}\; \tilde {\cal A}_0(\vec k,s) 
\end{eqnarray}
where the contour is to the right of all the singularities as in 
Eq.(\ref{antisqed}). As in the 
transverse case, at high temperatures the singularity structure is
dominated by the discontinuity across the cut that runs from $-ik$ to $+ik$
corresponding to Landau damping,  and the two plasmon poles at  
$\pm i\omega_0$. Deforming the   contour to pick up the cut
and pole contributions we get
\begin{eqnarray} 
{\cal A}_{0}(\vec k,t)={\cal A}_{0}^{pole}(\vec k,t)
+{\cal A}_{0}^{cut}(\vec k,t),\label{lpolecut} 
\end{eqnarray} 
where 
\begin{eqnarray} 
&&{\cal A}_{0}^{pole}(\vec k,t)=
-Z^l[T]\frac{\sin(\omega_0t)}{\omega_0}\, , \,
Z^l[T]=\left[\frac{\partial\Sigma^l(i\omega)}{\partial\omega^2}\right] 
_{\omega=\omega_o\simeq eT/3}^{-1} \label{zetalongi}  \nonumber\\
&&\mbox{and}  \!\!\!\!\!\!\!\!\!\!\!\!\!\!\!\!\!\!\!\!\!\!\!\! \\
&&{\cal A}_{0}^{cut}(\vec k,t)=
-\frac{2}{\pi}\int_0^kd\omega\frac{\Sigma^l_k(\omega)_I\;\sin(\omega t)} 
{\left[k^2+\Sigma^l_k(\omega)_R\right]^2+\left[\Sigma^l_k(\omega)_I
\right]^2} \label{lcutonly} \nonumber
\end{eqnarray}
Unlike the transverse components for which the sum rule is a
consequence of the
canonical commutation relations, for the longitudinal component there
is no equivalent sum rule because the field $A_0(x)$ is a non-propagating
Lagrange multiplier.

The long time, asymptotic behaviour of the longitudinal condensate is
dominated by the end-points of the integral (\ref{lcutonly}). The
end-point $ \omega = 0 $ yields contributions that vanish for long $ t
$ faster than any negative power of $ t $, as it was the case for the
transverse part (\ref{tcutonly}). We find that the  long time asymptotics
 for $t>>1/k $ is dominated by  the  end-point contribution at 
$ \omega = k $\cite{htl},
\begin{eqnarray} \label{asiL} 
&&{\cal A}_0^{cut}(\vec k,t)=
a_{asy}^{cut}(k,t)\left[1+{\cal
O}\left(\frac{1}{t}\right)\right] \;\;, \\
&&a_{asy}^{cut}(k,t)\equiv -{12 \over {e^2T^2}}\!\!
\int_0^{\infty}e^{-x}dx  {{\cos[ k t + \alpha(x,t) ]}
\over {\sqrt{\log^4{{cx}\over {kt}} + {{5\pi^2}\over 2} 
\log^2{{cx}\over {kt}} + {{9 \pi^4}\over {16}} }}} \cr \cr
&&\alpha(x,t) \equiv  \arctan{{\pi \log{{cx}\over {kt}}}\over 
{ {{3\pi^2}\over 4} + \log^2{{cx}\over {kt}}}}  \; , \;
c   \equiv  \frac12  \exp[ 2 + {{6k^2}\over {e^2T^2}}] \;.\nonumber
\end{eqnarray}

The integral in Eq.(\ref{asiL}) cannot be expressed in terms of
elementary functions and it is related to the $ \nu(x) $ function
\cite{erde}. For large $ t $, one can derive an asymptotic expansion
in inverse powers of $ \log [kt/c] $ by integrating Eq.(\ref{asiL}) by parts:
\begin{eqnarray}
&&a_{asy}^{cut}(k,t) \buildrel{t \to \infty}\over= 
-{12 \over {e^2T^2\; t}}\; {{\cos(kt + \beta(kt))}\over{{\log^2{{kt}\over
c} +  {{\pi^2}\over 4} }}} \left[1+{\cal
O}\left(\frac{1}{\log t}\right)\right] \nonumber \\
&&\beta(kt)  \equiv   -\arctan{{\pi \log{kt}}\over 
{\log^2{kt} -{{\pi^2}\over 4}}} \; ,\label{asiLD}
\end{eqnarray}

This expansion is not very good quantitatively unless $ t $ is very
large. For example, for $ kt = 500 ,\;  a_{asy}^{cut}(
k,t) $ given by Eq.(\ref{asiL})
approximates $ {\cal A}_0^{cut}(\vec k,t) $ up to $ 0.1 \% $,
whereas, the dominant term in Eq.(\ref{asiLD}) is about $ 30\% $
smaller than $ a_{asy}^{cut}(k,t) $\cite{htl}. 

To summarize, we gather
the final results for the asymptotic real-time evolution of the transverse
and longitudinal condensates in the linear approximation
{\bf (a) Transverse (no external source)}: 
\begin{eqnarray}
&&{\cal A}_{Ti}(\vec k,t) =   {\cal A}_{Ti}(\vec k,0) \left[
Z^t[T] \cos\omega_pt - \frac{12}{e^2T^2}\frac{\cos kt }{t^2}
\right] \nonumber \\
 && +  \dot{\cal A}_{Ti}(\vec k,0) \left[
Z^t[T] \frac{\sin\omega_pt}{\omega_p} - 
\frac{12}{e^2T^2}\frac{\sin kt }{kt^2}\right]
+{\cal O}\left(\frac{1}{t^3}\right) \nonumber
\end{eqnarray}
The sum rule (\ref{sumrule}) implies that the coherent field configuration
relaxes to an  asymptotic amplitude which is {\em smaller} than the initial, 
and the ratio of the final to the initial amplitude is completely determined by
the thermal wave function renormalization.     

{\bf (b) Longitudinal (impulsive external source): }
$$
{\cal A}_{0}(\vec k,t) =   
-Z^l[T]\;\frac{\sin(\omega_0t)}{\omega_0}+
 a_{asy}^{cut}(k,t)\left[ 1 + {\cal O}\left(\frac{1}{t}\right)\right].
$$
The transverse and longitudinal wave function renormalizations $Z^t[T]\;,
Z^l[T]$ are given by Eqs. (\ref{zetatrans}) and (\ref{zetalongi}) respectively
and $a_{asy}^{cut}(k,t)$ is given by Eqs. (\ref{asiL})-(\ref{asiLD}) while
the positions of the poles $\omega_p  ,\; \omega_0$ can be obtained by the
finding the zeroes of the denominators using eqs.(\ref{tbelowcut}) 
and (\ref{lbelowcut}), respectively.

In summary, we find that the long time dynamics is dominated by the Landau
damping thresholds at $\omega = \pm k$, not by the $\omega \approx 0$ region of
the spectral density. The early time dynamics is determined by moments of the
total spectral density.

While an understanding of the real-time relaxation of non-equilibrium, 
inhomogeneous field configurations is of fundamental importance in the  physics
of relaxation in the QGP \cite{blaizotqgp},   
such a calculation also has  phenomenological implications 
for sphaleron induced B-violating processes.

One of our main focus is to assess in detail  
 the real time non-equilibrium dynamics of soft excitations in the
 plasma with particular attention to a critical analysis of the
 long-standing 
belief that the small frequency region of the spectral function dominates
the long time relaxational dynamics. We find, to the contrary that the
Landau damping {\bf thresholds} at $\omega = \pm k$ determine the long-
time dynamics and that the early time dynamics is sensitive to several
 moments of the total spectral density. 
In ref.\cite{htl} we concentrated on the case of scalar 
electrodynamics (SQED) since this theory has the same HTL structure (to lowest
order) as the non-abelian case \cite{lebellac,rebhan}. Most of our results can
therefore, be taken over to the non-Abelian case  with little or no
changes at least to the leading order in HTL.  

When finite temperature infrared divergences are present, like for
scalar condensates 
coupled with gauge fields, the amplitude approximation
 combined with the resummation of the infrared
divergences by dynamical renormalization group leads to anomalous
logarithmic relaxation and logarithmic phases in time \cite{infra}.

\section{Non-linear Quantum Evolution in Scalar Field Theories}

Our goal is to deal with the out of equilibrium evolution for {\bf
large} energy densities in field theory.   That is, a large number of
particles per 
volume $ m^{-3} $, where $ m $ is the typical mass scale in the
theory.   The most familiar techniques of field theory, based on the S-matrix
formulation of transition amplitudes and perturbation theory apply in
the opposite limit of low 
energy density and since they only provide information on {\em in}
$\rightarrow$ {\em out} matrix elements, are unsuitable for
calculations of time dependent expectation values.

Similar tools are also necessary to describe consistently the dynamical
processes in the Early Universe\cite{eri97,nos2}. In particular it has
been recognized that 
novel phenomena associated with parametric amplification of quantum 
fluctuations can play an important role in the process of reheating and
thermalization\cite{eri97,nos2,cosmo}.  It must be noticed
 that the dynamics in cosmological spacetimes is
dramatically different to the dynamics in  Minkowski spacetime. 
Both in fixed FRW and de Sitter\cite{cosmo} backgrounds and in a
dynamical geometry\cite{din} the dynamical evolution is qualitatively and
quantitatively different to the Minkowski case considered in the present
paper.

Our program to study the dynamical aspects of relaxation out
of equilibrium both in the linear and non-linear regime has revealed
new features of relaxation in the collisionless regime in scalar
field theories\cite{nos2,inhomo}. Recent investigation of scalar
field theories in the non-linear regime, including self-consistently
the effects of quantum backreaction in an energy conserving and 
 renormalizable framework have pointed out to a wealth of interesting
non-perturbative phenomena both in the broken and unbroken symmetry
phases\cite{nos2}. These new phenomena are a consequence of the
non-equilibrium 
evolution of an initial state of large energy density which results
in copious particle production  leading to non-thermal and
non-perturbative distribution of particles. Our studies have focused
on the situation in which the {\em 
amplitude} of the expectation value of the scalar field is
non-perturbatively large, 
$A \approx \sqrt{\lambda} <\Phi> /m \approx {\cal O}(1)$ ($m$ is the
mass of the scalar field and $\lambda$ the self-coupling) and most of
the energy of the initial 
state is stored in the `zero mode', i.e. the (translational
invariant) expectation value of 
the scalar field $\Phi$. Under these circumstances the initial energy
density $\varepsilon 
\approx m^4/\lambda$. During the dynamical evolution the energy
initially stored in one 
(or few) modes of the field is transferred to other modes resulting in
copious particle production initially either by parametric
amplification of quantum fluctuations in the 
unbroken symmetry phase, or spinodal instabilities in the broken
symmetry phase. Furthermore, we have considered in ref.\cite{tsu} the
evolution of an initial state with a large number of quanta with
a distribution around a momentum $ |\vec k_0| $ corresponding to a  
 {\em thin spherical shell} in momentum space, a `tsunami'.

In both cases the  
mechanism of energy dissipation and particle production results in a
number of produced 
particles per unit volume ${\cal N} \propto m^3/\lambda$ which for
weak coupling is non-perturbatively large\cite{nos2,tsu}.   We call
`linear regime' to this first stage dominated either by parametric or
spinodal unstabilities.

We recognized\cite{nos2} a new {\em dynamical} time
scale $ t_1 $ where the linear regime ends. By the time  $ t_1 $
the effects of the quantum fluctuation on the
dynamical evolution become of the same order as the classical contribution
given by the evolution of the expectation value of the field.
The `non-linear' regime starts by the time $ t_1 $.
In the case of broken symmetry, this time scale corresponds to the
spinodal scale at which the backreaction of quantum fluctuations
shut-off the spinodal instabilities.  At this
scale non-perturbative physics sets in and the non-linearities 
of the full quantum theory determine the evolution. This time scale $ t_1 $,
which we call the {\em non-linear time}, is a non-universal feature of
the dynamics and depends strongly on the initial state and
non-perturbatively on the coupling, as $ t_1 \propto
\log[\lambda^{-1}] $ for 
weak coupling\cite{nos2}. 

There are two very important  
parameters that influence the quantum dynamics: the strength of the
coupling constant $\lambda$ and the initial energy density in units
of the scalar field mass $m$. If in the initial state most of the
energy is stored in the zero mode, the energy density is determined by
the amplitude $ A $ of the order parameter. The value of this field
amplitude determines  
the regime of applicability of perturbation theory methods.  That is,
the amplitude approximation considered in secs. II and III fails to
catch important aspects of the physics when $ A $ is not small. 

We consider now  the {\bf non-perturbative} regime in 
which $ A =  \sqrt{ \lambda} \; <\Phi>/ m \simeq {\cal{O}}(1) $. It is
important to point out that for 
large field amplitude, even for very weakly coupled theories
non-linear effects will be important and must be treated
non-perturbatively. This is 
the case under consideration. Having recognized the non-perturbative
nature of the problem for large amplitudes we must invoke a
non-perturbative, consistent calculational scheme which respects the
symmetries (continuous global symmetries and energy-momentum
conservation), is renormalizable and lends itself to a numerical treatment. 

We are thus led to  consider  the $O(N)$ vector model with quartic
self-interaction\cite{nos2} and the scalar field in the vector  
representation of $O(N)$. 

The action and Lagrangian density are given by,
\begin{eqnarray}
S  &=&  \int d^4x\left\{\frac{1}{2}\left[\partial_{\mu}{\vec{\Phi}}(x)\right]^2
-V(\vec{\Phi}(x))\right\}\; ,\label{action} \cr \cr
V(\vec{\Phi})  &=&  \frac{\lambda}{8N}\left(\vec{\Phi}^2+\frac{2N
m^2}{\lambda}\right)^2 - {{N\;m^4}\over { 2 \lambda}} \quad . \label{potential}
\end{eqnarray}
We write the field $\vec{\Phi}$ as $\vec{\Phi} = (\sigma, \vec{\pi} )$
where $\vec{\pi}$ represents the 
$N-1$ `pions', and choose the coupling
 $\lambda$ to remain fixed in the large $N$ limit. Due to lack of
space we only considered here the unbroken symmetry case $ m^2 > 0
$. The broken symmetric case ($ m^2 < 0 $) is considered in ref.\cite{nos2}.

We can decompose the field $\sigma$ into its expectation value and fluctuations
$\chi( \vec{x},t )$ about it:
\begin{equation}
\sigma  (\vec{x},t ) = \Phi(t)\; \sqrt{N}+ \chi ( \vec{x},t) \; ,
\end{equation}
with $ \Phi(t) $ being a c-number of order one in the $ N \to \infty $
limit and $\chi$ an operator.

We shall not rederive here the field evolution equations for 
translationally invariant quantum states. In the leading order
in the large $ N $ approximation the theory becomes Gaussian at the
expense of a self-consistent condition\cite{nos2,cooper,eri96}, this in turn
entails that the  Heisenberg field operator $\Psi(\vec x,t)$ can be
written as 
$$
\Psi(\vec x , t) = \int {{d^3 k} \over {(2\pi)^3}}
\frac{1}{\sqrt{2 }}\left[  \;
a_{ \vec k} \; f_k(t) \; e^{i \vec k \cdot \vec x} + a^{\dagger}_{ \vec
k} \; f^*_k(t) \;  e^{-i \vec k \cdot \vec x}  \; \right],
$$
where $a_k \, , a^{\dagger}_k$ are the canonical creation and annihilation
operators, the mode functions $f_k(t)$ are solutions of the Heisenberg
 equations of motion\cite{nos2,cooper,eri96} given below
(\ref{modo0}-\ref{conds2}). 

Our choice of initial conditions on the density matrix is that of
the vacuum for the instantaneous modes of the Hamiltonian at the
initial time\cite{nos2,eri96}. Therefore we choose the initial
conditions on the mode functions to represent positive energy particle
states of the instantaneous Hamiltonian at $t=0$, which is taken to be
the initial time. That is, 
$$
f_k(0)= \frac{1}{\sqrt{W_k}} \; ; \; \dot{f}_k(0) = -i \sqrt{W_k} \; \; ; \; \;
W_k= \sqrt{k^2+M^2_0}\; \; ,  
$$
 where the mass $ M_0 $ is defined below.
With these boundary conditions, the mode functions
$f_k(0)$ correspond to positive frequency modes (particles) of the
instantaneous quadratic Hamiltonian for oscillators of mass $ M_0 $.

We point out that the behavior of the system depends mildly on the
initial conditions on the mode functions as we have found by varying
them within a wide range. In 
particular, the various types of linear and nonlinear resonances 
are independent of these initial conditions\cite{nos2,eri96}.

We introduce the following dimensionless quantities:
\begin{eqnarray}
&& \tau = |m|\, t \;  , \;   q = \frac{k}{|m| }
\;  , \;  \Omega_q= \frac{W_k}{|m|} \; \; , \;
\eta^2(\tau) = \frac{\lambda }{2 |m|^2 }\; \Phi^2(t)  \; \; , \nonumber
\\ 
&& g \Sigma(\tau) = 
\frac{\lambda}{2|m|^2 } \left[ \langle \Psi^2(t) \rangle_R- \langle
 \Psi^2(0) \rangle_R  
\right]  \; \; , \; \; \left( \; \Sigma(0) = 0  \;\right) \nonumber \\
&& g = \frac{\lambda}{8\pi^2}  \; \; , \; \;
\varphi_q(\tau) \equiv \sqrt{|m|} \; f_k(t)  \; \; .\nonumber
\end{eqnarray}
Here  $ \langle \Psi^2(t) \rangle_R $ stands for the renormalized
composite operator [see eq.(\ref{sigmafin}) for an explicit expression].

\subsection{Evolution equations in the large $N$ limit}

In terms of the dimensionless variables
introduced above  the renormalized equations of motion are found to be
(see references\cite{nos2,eri96}):

\begin{eqnarray}
& & \ddot{\eta}+ \eta+
\eta^3+ g \;\eta(\tau)\, \Sigma(\tau)  = 0 \label{modo0} \\
& & \left[\;\frac{d^2}{d\tau^2}+q^2+1+
\;\eta(\tau)^2 + g\;  \Sigma(\tau)\;\right]
 \varphi_q(\tau) =0 \; , \label{modok}  \\
&& \varphi_q(0) = {1 \over {\sqrt{ \Omega_q}}} \, , \,
{\dot \varphi}_q(0) = - i \; \sqrt{ \Omega_q} \, , \,
\eta(0) = \eta_0  \, , \,{\dot\eta}(0) = 0 \label{conds2}
\end{eqnarray}
Hence, ${\cal M}^2(\tau) \equiv 1+\;\eta(\tau)^2 + g\;  \Sigma(\tau) $
plays the r\^ole of a (time dependent)   effective mass squared.  

We will choose $ \Omega_q $
such that at $t=0$ the quantum fluctuations are in the ground
state of the oscillators at the initial time. Since
$g\Sigma(0)=0$, we choose $ M_0 \equiv \sqrt{1 + \eta_0^2} $.
$g \Sigma(\tau)$ is given by the self-consistent
condition\cite{nos2,eri96} 
\begin{eqnarray}
g \Sigma(\tau) & = & g \int_0^{\infty} q^2
dq \left\{ 
\mid \varphi_q(\tau) \mid^2 
-\frac{1}{\Omega_q}   \right. \nonumber \\
 &   & \left. + \frac{\theta(q-1)}{2q^3}\left[ 
 -\eta^2_0 + \eta^2(\tau) + g \; \Sigma(\tau) \right] \right\} \;
. \label{sigmafin} 
\end{eqnarray}
We thus see that the effective mass at time $ \tau $ contains all
$q$-modes and the zero mode at the same time $ \tau $. The evolution
equations are then nonlinear but local in time in the infinite $N$ limit.

\subsection{The early time evolution: parametric resonance}

In the weak coupling regime the back-reaction term $ g
\Sigma(\tau) $ is small for small $ g $ during an interval say $ 0
\leq \tau < \tau_1 $\cite{nos2,eri96}. This time $\tau_1$ will 
be called the nonlinear time and it determines the time scale when the
backreaction  effects and therefore the quantum fluctuations and
non-linearities become important.  

For times $ \tau < \tau_1 , \;  g \Sigma(\tau) $
can be neglected  and then eq.(\ref{modo0}) reduces to the classical
equation of  motion 
\begin{equation}
 \ddot{\eta}+ \eta+ \eta^3  = 0 \; . \label{classical}
\end{equation}
The solution of this equation with the initial conditions (\ref{conds2})
can be written  in terms of elliptic functions 
$$
\eta(\tau) = \eta_0\; \mbox{cn}\left(\tau\sqrt{1+\eta_0^2},k\right)
\quad , \quad
k = {{\eta_0}\over{\sqrt{2( 1 +  \eta_0^2)}}}\; , 
$$
where cn stands for the Jacobi cosine.  

Inserting this form for 
$\eta(\tau)$ in eq.(\ref{modok}) and again neglecting $ g \Sigma(\tau) $ yields
\begin{equation}\label{modsn}
 \left[\;\frac{d^2}{d\tau^2}+q^2+1+  \eta_0^2\;
\mbox{cn}^2\left(\tau\sqrt{1+\eta_0^2},k\right) \;\right]
 \varphi_q(\tau) =0 \; .
\end{equation}

This is the Lam\'e equation for a particular value of the coefficients that
make it solvable in terms of Jacobi functions \cite{nos2,eri96}.  

Since the coefficients of eq.(\ref{modsn}) are periodic with period $ 2 \omega
$ (notice that $\eta(\tau + 2 \omega ) =  - \eta(\tau)$),
the mode functions can be chosen to be quasi-periodic (Floquet type) with
quasi-period $ 2 \omega $,
\begin{equation}\label{floq}
 U_q(\tau + 2  \omega) =   e^{i F(q)} \; U_q(\tau),
\end{equation}
where the Floquet indices $ F(q) $ are independent of $\tau$.  In the allowed
zones, $ F(q) $ is  real  and the functions $  U_q(\tau) $ 
are bounded. In the forbidden zones $ F(q) $ has a non-zero imaginary
part and the amplitude of the solution either grows or decreases
exponentially.

It turns out in the present case that there is only one forbidden band
for $ q^2 > 0 $ running in the interval $ 0 < q^2 < \eta_0^2/2 $.

The modes from the forbidden band $ 0 < q <
 \eta_0/\sqrt2 $ dominate $ \Sigma(\tau) $. For $ 0 < \tau < \tau_1 $,
  $ \Sigma(\tau) $ oscillates with an exponentially growing
 amplitude. This amplitude (envelope)  $ \Sigma_{env}(\tau) $
can be represented to a very good approximation by the formula\cite{nos2,eri96}
\begin{equation}\label{polenta}
 \Sigma_{env}(\tau) = { 1 \over { N \, \sqrt{\tau}}}\; e^{B\,\tau}\; ,
\end{equation}
where $ B $ and $ N $ are functions of $ \eta_0 $ given by
\begin{eqnarray}\label{ByN}
B(\eta_0) &=&  \displaystyle{
8\, \sqrt{1+\eta_0^2}\; {\hat q } \; (1 - 4 {\hat q }) +  O(
{\hat q }^3) }\; , \cr \cr
N(\eta_0)  &=& {4 \over {\sqrt{ \pi}}} \; \sqrt{  {\hat q }}\;
{{ ( 4 + 3 \, \eta_0^2) \, \sqrt{  4 + 5 \, \eta_0^2}}\over{
 \eta_0^3 \, (1+\eta_0^2)^{3/4}}} \left[ 1 + O ({\hat q })\right]\; \; .
\end{eqnarray}
and the elliptic nome $ {\hat q } $ can be written as a function of $
\eta_0 $ as  
\begin{equation}\label{qaprox}
{ \hat q }(\eta_0) =  \frac12 \;  {{ (1+\eta_0^2)^{1/4} 
-  (1+\eta_0^2/2)^{1/4}}
\over { (1+\eta_0^2)^{1/4} +  (1+\eta_0^2/2)^{1/4}}}  \; .
\end{equation}
 with an error smaller than $\sim 10^{-7} $.

Using this estimate for the quantum fluctuations $ \Sigma(\tau) $, we
estimated  the nonlinear time scale $\tau_1$ at which the back-reaction becomes
comparable to the classical terms. Such a time is defined by
$ g \Sigma(\tau_1) \sim (1+\eta^2_0/2) $,\cite{nos2}
$$
\tau_1 \approx {1 \over B(\eta_0)} 
\, \log{N(\eta_0)\,(1+\eta^2_0/2) \over { g \,\sqrt{
B(\eta_0)}}}\; .
$$
 \begin{figure}[t] 
\epsfig{file=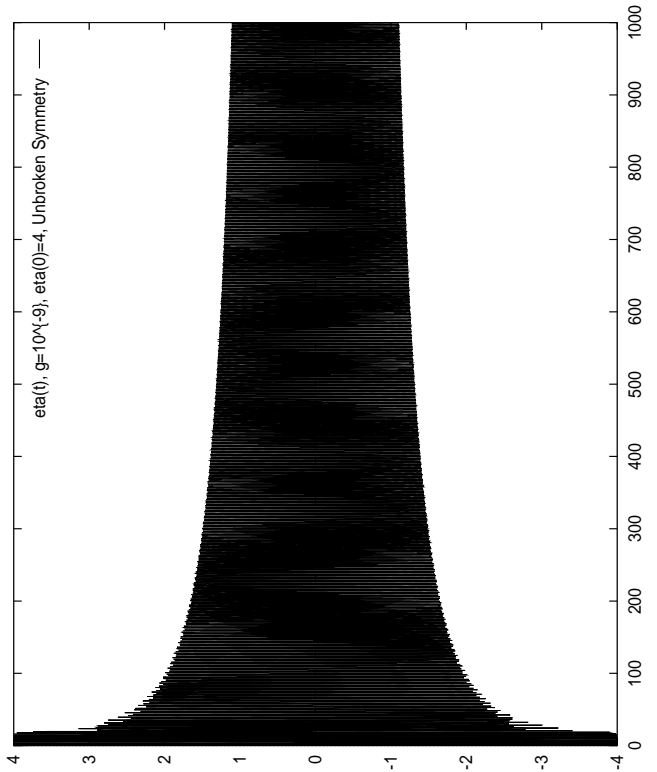,width=8.5cm,height=11cm} 
\caption{ The zero mode $\eta(\tau)$ vs. $\tau$ for the unbroken
symmetry case with $ \eta_0=4 $, $ g=10^{-9} $.
\label{fig1}} 
\end{figure} 
The time interval from $\tau=0$ to $\tau\sim \tau_1$ is when most of
the particle production takes place. Around $\tau \sim \tau_1 $ the quantum
fluctuation become large enough to begin shutting-off the growth of the modes
and particle production slows down dramatically. This dynamical time
scale separates two distinct types of dynamics, for $\tau < \tau_1$ the 
evolution of the quantum modes $ \varphi_q(\tau) $ is essentially linear, 
the backreaction effects are small and particle production proceed via
parametric amplification.  For $ \tau > \tau_1 $ the quantum backreaction
effects are as important as the tree level term $ \eta(\tau)^2 $ and
the dynamics is fully  non-linear.   

The growth of the unstable modes in the forbidden band shows that
particles are created copiously ($\sim 1/g $ for $ \tau \sim
\tau_1 $). Initially, ($\tau = 0$) all the energy is in the classical zero
mode (expectation value). Part of this energy is rapidly transformed
into particles 
through parametric resonance during the interval  $ 0 < \tau < \tau_1$.
At the same time, the field expectation value decreases as
is clearly displayed in  fig. 1. 

The momentum distribution of the  particles produced by parametric resonance
follows the Floquet index and is peaked at $ q \approx \frac12 \eta_0
\,( 1 - {\hat q})$\cite{nos2,eri96}. This is shown in fig. 2. 

\begin{figure}[t] 
\epsfig{file=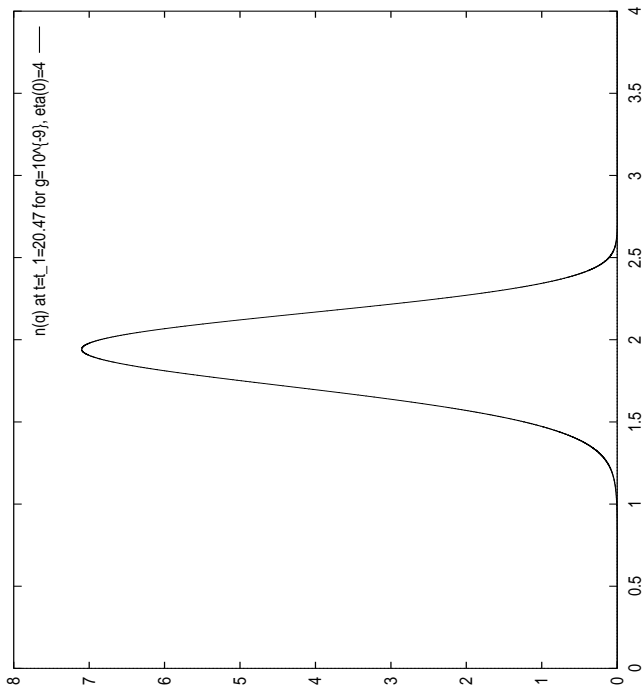,width=8.5cm,height=10cm} 
\caption{ Momentum distribution of the produced particles 
$ n(q) \equiv q^2 \, N^{ad}_q(\tau) $ at the nonlinear
time $ \tau = \tau_1 $ for $ \eta_0=4 $, $ g=10^{-9} $.
\label{fig2}} 
\end{figure} 
For late times, typically larger than 3 or 4 times $ \tau_1 $, the
effective mass  tends asymptotically to a constant value 
$ {\cal M}^2_{\infty} = 1 + {{\eta_0^2}\over 2} $. This fact implies that the
modes becomes effectively {\bf free}.

At the same time, the field expectation
value relaxes to zero asymptotically with 
a non-universal power law. The initial energy density which
is non-perturbatively large goes completely into  production of
massive particles.  The asymptotic particle distribution
is localized within a  band  determined by the initial conditions with
non-perturbatively large amplitude $ \sim 1/\sqrt{\lambda}$ which
could be described as a `semiclassical condensate' in the unbroken
phase [see fig. 3]. 

\begin{figure}[t] 
\epsfig{file=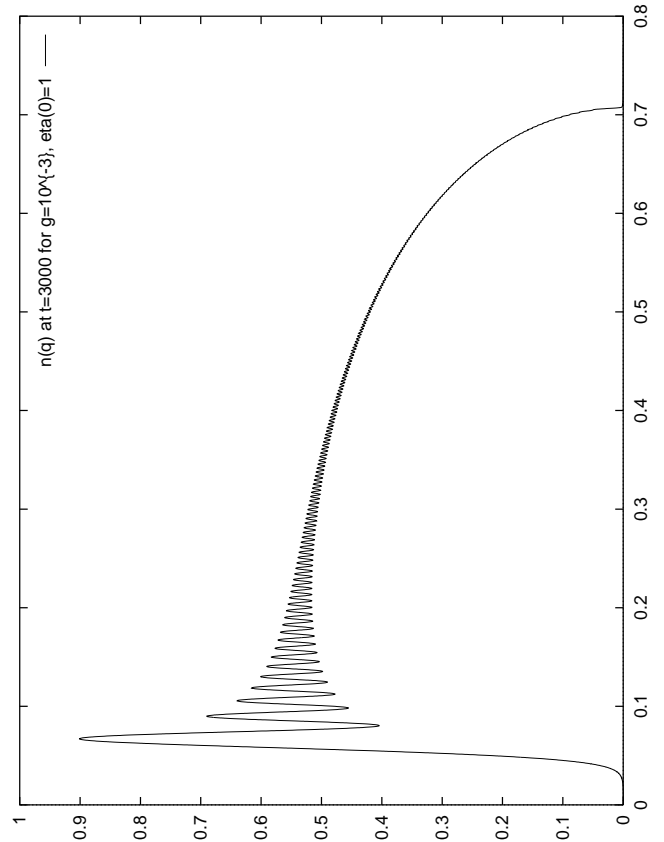,width=8.5cm,height=12cm} 
\caption{Momentum distribution of the produced particles at $ \tau = 3000 $ for
$ \eta_0=1  , \, g=10^{-3} $. Notice the main peak at $ q=0.067 $, in
good agreement with the estimate $ K_1/ \tau $ given in ref.[2].
\label{fig3}}
\end{figure} 

 The novel result that emerged   combinining
 numerical methods and dynamical renormalization group analysis is the presence
of {\em non-linear resonances} that lead to  asymptotic  relaxation
 described by {\em non-universal power laws}\cite{nos2}. These power laws are
 determined by {\em dynamical anomalous exponents} which depend
 non-perturbatively  on the coupling.

\section{acknowledgements}
D. B. thanks the N.S.F. for partial support through grant PHY-9605186.
R. H. and S. P. K. were supported by DOE grant DE-FG02-91-ER40682.
The authors acknowledge support from NATO. H. J. de V. thanks the
organizers of TFT98 where this work has been  reported.
 
\end{document}